\documentclass[sigconf]{acmart}
\AtBeginDocument{%
  \providecommand\BibTeX{{%
    \normalfont B\kern-0.5em{\scshape i\kern-0.25em b}\kern-0.8em\TeX}}}

\usepackage{pifont}
\newcommand{\cmark}{\ding{51}}%
\newcommand{\xmark}{\ding{53}}%

\usepackage{tabularx}
\usepackage{multirow}
\usepackage{dcolumn} 
\newcolumntype{d}[1]{D{.}{.}{#1}}
\usepackage{color, colortbl}

\definecolor{Gray}{gray}{0.95}

\copyrightyear{2024}
\acmYear{2024}
\setcopyright{rightsretained}
\acmConference[UIST '24]{The 37th Annual ACM Symposium on User Interface Software and Technology}{October 13--16, 2024}{Pittsburgh, PA, USA} \acmBooktitle{The 37th Annual ACM Symposium on User Interface Software and Technology (UIST '24), October 13--16, 2024, Pittsburgh, PA, USA} \acmDOI{10.1145/3654777.3676455}
   
\acmISBN{979-8-4007-0628-8/24/10}

\definecolor{darkgreen}{rgb}{0, 0.5, 0}
\definecolor{darkred}{rgb}{0.8, 0, 0}
\newcommand{\greencmark}{{\color{darkgreen}\cmark}}
\newcommand{\redxmark}{{\color{darkred}\xmark}}

\usepackage{soul}
\usepackage{enumitem}

\begin{document}
\setlist[itemize]{leftmargin=*}

\title{EgoTouch: On-Body Touch Input Using AR/VR Headset Cameras}

\author{Vimal Mollyn}
\affiliation{
 \institution{Carnegie Mellon University}
 \city{Pittsburgh}
 \state{PA}
 \country{USA}}
\email{vmollyn@andrew.cmu.edu}

\author{Chris Harrison}
\affiliation{
 \institution{Carnegie Mellon University}
 \city{Pittsburgh}
 \state{PA}
 \country{USA}}
\email{chris.harrison@cs.cmu.edu}

\renewcommand{\shortauthors}{Mollyn and Harrison}

\begin{abstract}
In augmented and virtual reality (AR/VR) experiences, a user’s arms and hands can provide a convenient and tactile surface for touch input. Prior work has shown on-body input to have significant speed, accuracy, and ergonomic benefits over in-air interfaces, which are common today. In this work, we demonstrate high accuracy, bare hands (i.e., no special instrumentation of the user) skin input using just an RGB camera, like those already integrated into all modern XR headsets. Our results show this approach can be accurate, and robust across diverse lighting conditions, skin tones, and body motion (e.g., input while walking). Finally, our pipeline also provides rich input metadata including touch force, finger identification, angle of attack, and rotation. We believe these are the requisite technical ingredients to more fully unlock on-skin interfaces that have been well motivated in the HCI literature but have lacked robust and practical methods.


\end{abstract}


\begin{teaserfigure}
  \includegraphics[width=\textwidth]{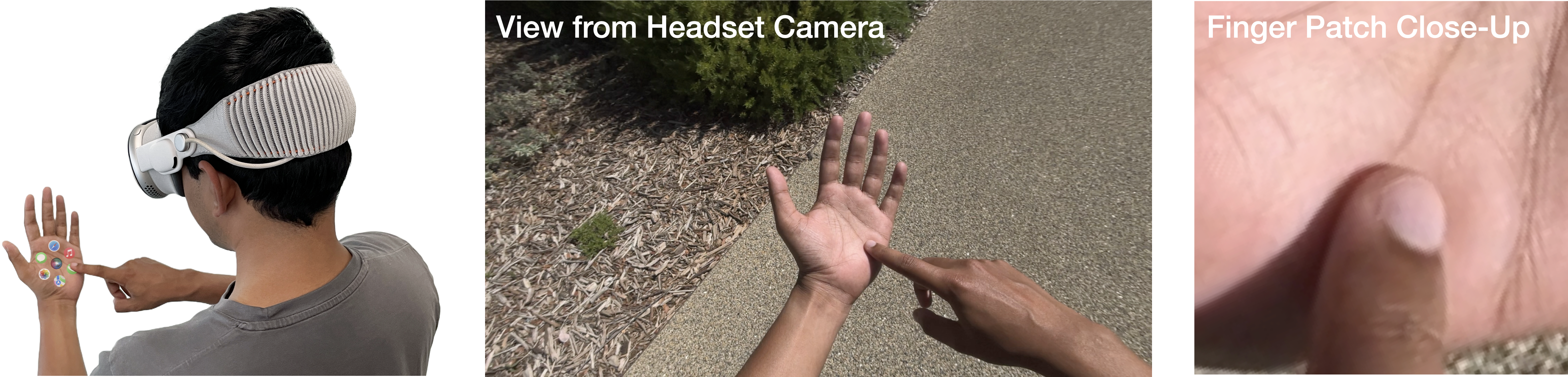}
  \caption{Using RGB cameras already present in modern XR headsets for passthrough, EgoTouch can enable on-body touch input with no additional accessories or sensors.}
  \label{fig:teaser}
\end{teaserfigure}

\maketitle

\begin{table*}[t]
\centering
\resizebox{\textwidth}{!}{
\begin{tabular}{l c c c c c c c c c c c}
\toprule
\textbf{\begin{tabular}[l]{@{}l@{}}System \\ Name \end{tabular}} & \textbf{\begin{tabular}[c]{@{}c@{}}Camera \\ Sensor \end{tabular}} & \textbf{\begin{tabular}[c]{@{}c@{}}Supports \\ Skin  Input\end{tabular}} & \textbf{\begin{tabular}[c]{@{}c@{}}Uninstrumented \\ Inputting Arm \end{tabular}} & \textbf{\begin{tabular}[c]{@{}c@{}}Uninstrumented \\ Receiving Arm/Skin \end{tabular}} & \textbf{\begin{tabular}[c]{@{}c@{}}Sensor \\ Location\end{tabular}} & \textbf{\begin{tabular}[c]{@{}c@{}}Touch \\ Force\end{tabular}} & \textbf{\begin{tabular}[c]{@{}c@{}}Calibration \\ Free\end{tabular}} & \textbf{\begin{tabular}[c]{@{}c@{}}Worn \\ Sensor\end{tabular}} & \textbf{\begin{tabular}[c]{@{}c@{}}Study \\ Conditions\end{tabular}} & \textbf{\begin{tabular}[c]{@{}c@{}}Click Event \\ Accuracy \end{tabular}} \\

\midrule

ShadowTouch~\cite{shadowtouch} & RGB & \redxmark & \redxmark & \greencmark & Wrist & \redxmark & \greencmark & \redxmark & Sitting, Indoor & 99.1\% \\
DIRECT~\cite{DIRECT} & Depth \& IR & \redxmark & \greencmark & \greencmark & Ceiling & \redxmark & \redxmark & \redxmark & Indoor & 99.3\% \\
FarOut Touch~\cite{farouttouch} & Depth & \redxmark & \greencmark & \greencmark & Tripod & \redxmark & \greencmark & \redxmark & Standing, Indoor & 96.9\% \\ 
TapLight~\cite{StructuredLightSpeckle} & IR & \redxmark & \greencmark & \greencmark & Headset & \redxmark & \greencmark & \greencmark & Sitting, Indoor & 95.3\% \\ 
OmniTouch~\cite{omnitouch} & Depth & \greencmark & \greencmark & \greencmark & Shoulder & \redxmark & \greencmark & \greencmark & Standing, Indoor & 96.5\% \\
WatchSense~\cite{watchsense} & Depth & \greencmark & \greencmark & \redxmark & Forearm & \redxmark & \greencmark & \greencmark & Sitting \& Standing, Indoor & 77.0\% \\
TriPad~\cite{dupre:hal-04497640tripad} & (Hand tracking API) & \redxmark & \greencmark & \greencmark & Headset & \redxmark & \greencmark & \greencmark & Sitting, Indoor & 99.0\% \\
PressureVision++~\cite{PressureVisionPlusPlus} & RGB & \redxmark & \greencmark & \greencmark & Table & \greencmark & \greencmark & \redxmark & Sitting, Indoor & 89.3\% \\

\rowcolor{Gray}
EgoTouch (ours) & RGB & \greencmark & \greencmark & \greencmark & Headset & \greencmark & \greencmark & \greencmark & Standing \& Walking, Indoor \& Outdoor & 95.6\% \\

\bottomrule
\end{tabular}}
\vspace*{2mm}\caption{Comparison of highly-related, vision-based touch sensing systems. EgoTouch is the only on-body touch system that does not instrument the user's arms or skin, is calibration-free, and works across diverse and dynamic environments. } 
\label{tab:touch-systems-overview}
\end{table*}

\section{Introduction}

In augmented and virtual reality (AR/VR) experiences, a user's arms and hands can serve as convenient and tactile input surfaces, which have been shown to be faster~\cite{wang2015palmtype, gustafson2013understandingfast}, more precise~\cite{lin2011pubpreciseonskin}, and more comfortable~\cite{weigel2014more} than in-air input. The lack of tactility in today's "floating" AR/VR interfaces is perhaps best exemplified by the large virtual keyboards in both Apple's Vision Pro and Meta's Quest series of headsets. Despite their large size, interacting on these keyboards is notably worse than even a diminutive smartphone keyboard. We believe bringing practical and robust on-skin touch tracking to AR/VR headsets can retain the convenience and flexibility of operating in a bare-hands manner (vs. controllers), while simultaneously offering rich and useful haptics, including proprioceptive cues, to further aid performance. 

For these reasons, technical approaches to enable on-body input have been studied for more than a decade (which we review in the next section). Unfortunately, few technologies meet three pillars we believe are essential for consumer adoption. 1) No user instrumentation (i.e., users wear only the headset with no other accessories). 2) Works across users/sessions/environments with no calibration. 3) Real-time implementation that runs on mobile hardware. 

Our system, EgoTouch, achieves these three criteria. Moreover, it does this using only sensors already present in modern headsets: RGB cameras used for pass-through and hand tracking. EgoTouch exposes a conventional event state machine, including both touch-down and touch-up events (many systems can only provide touch-down events). We also expose additional touch metadata to end-user applications, most notably touch force, which other systems cannot provide. At the heart of our system is an optimized deep learning model, with an inference time of less than 0.6~ms on an Apple M2 processor (like that found in the Apple Vision Pro), meaning our model can run at 90 FPS or higher with minimal impact on application performance.

\begin{figure}[b]
    \centering
    \includegraphics[width=\linewidth]{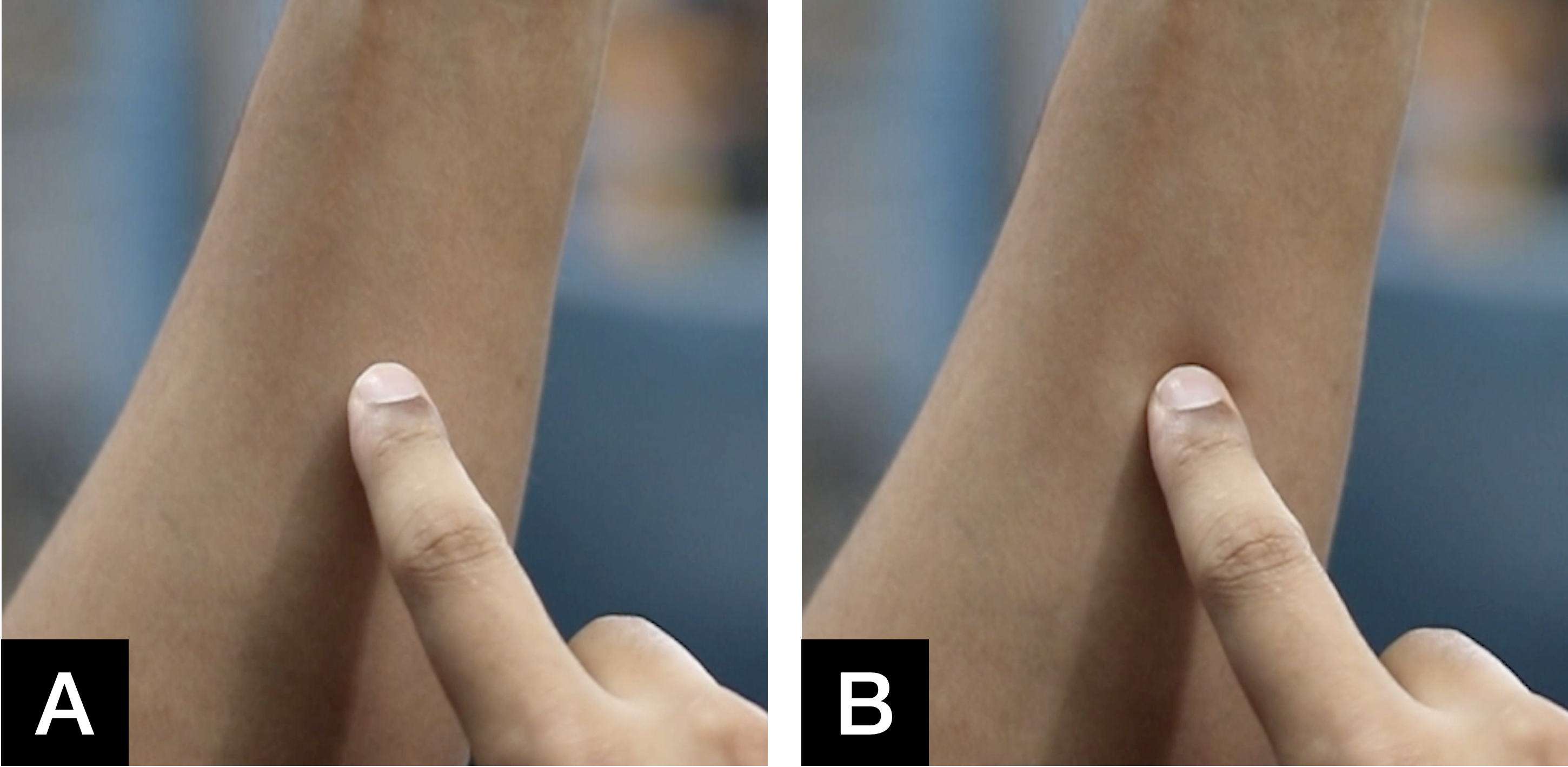}
    \caption{A finger hovering just above the skin (A) vs. touching the skin (B). Note the deformation of the host skin interacts with the ambient lighting, characteristically altering the shadows and shading.}
    \label{fig:touch_effect}
\end{figure}

The core insight that prompted our computer-vision-driven approach was that the skin deforms slightly in response to touch contact. This characteristically alters the local shading of the proximate host skin, as well as the shadow cast by the finger. An example of this effect is shown in Figure~\ref{fig:touch_effect} and Video Figure. Remarkably, this works across a wide range of lighting conditions (examples in \autoref{fig:lighting-conditions}), and works for surprisingly subtle touches. We encourage readers to tap the palm of their hand to see this effect firsthand. 

In addition to our system contribution, we endeavored to run a best-in-class user study, moving beyond prior work in rigor and ecological validity. Notably, this included collecting data while participants were both still and walking, in bright and dark lighting conditions, and across diverse skin tones, including the darkest complexions on the Fitzpatrick Phototyping Scale \cite{Fitzpatrick1975}. Overall, our system had a 96.4\% true positive touch detection rate with a 5.6\% false positive rate. Mean absolute touch force error was just 6.8\%. If we correct for synchronization error between our ground truth and RGB camera, accuracy further improves to 97.6\% true positive touch detection and false positive rate further reduces to 2.6\% (note that in a real commercial system, there would be no mismatch in sensor streams as there is no ground truth sensor). Finally, we believe a larger corpus of training data would also help unlock commercial-level accuracies.

\section{Related Work}
Our work was inspired by over a decade of interest in detecting ad-hoc touches on various surfaces, including the skin. We first review prior work that instruments the arm that is driving input. Next, we look at systems that instrument the arm that receives input. We then cover systems that instrument both the inputting arm and receiving arm. Finally, we review highly related camera-based touch sensing methods, including systems that do not instrument the users' arms, which are the most related prior work. For an in-depth review of the on-skin touch sensing literature, we refer readers to Bergström and Hornbæk's excellent survey \cite{HCIOnTheSkinSurvey}.

\subsection{Instrumented Inputting Arm}
A common approach for detecting touches to a user's skin is to instrument the hand driving the input. "Ready, Steady, Touch"~\cite{Shi2020ReadySteadytouch} attached an IMU to the top of the fingernail and was able to detect touches to the arm at 92.7\% and to the back of the hand at 90.4\%. TouchCam~\cite{stearns2018touchcam} used a combination of finger-mounted IR sensors, IMU, and camera to detect various touch locations on the body; touch events were segmented using a threshold from finger-mounted IR sensors. STAR~\cite{taejunSTAR} wrapped capacitive tape around the thumbs to detect taps. ElectroRing~\cite{Kienzle2021ElectroRingSP} used a ring form factor to inject high-frequency alternating current through the body; touches could be detected by measuring impedance changes from skin contact. Touch accuracy was 99.4\%, but required a per-user threshold calibration. Z-Ring~\cite{zring} used active electrical field sensing from a ring to detect a variety of single-handed and two-handed gestures, including taps to the skin, with an accuracy of 86.2\% with a user-independent model. More similar to our work, ShadowTouch~\cite{shadowtouch} used a forward-facing light source, worn at the wrist, to generate distinct shadow features that could be detected on a table by a desk-mounted camera with 99.1\% accuracy. A commonality of the above methods is they require instrumenting the user with an accessory, which adds cost and complexity. 

\subsection{Instrumented Input-Receiving Arm}
Prior work has also explored instrumenting the input-receiving skin surface with various sensors to enable touch sensing capabilities. Starting most simple, DisplaySkin~\cite{displayskin} simply wrapped a flexible touch screen around the user's wrist. iSkin \cite{iskin}, Multi-touch-skin~\cite{multitouchskin}, and DuoSkin~\cite{duoskin} created flexible capacitive touch sensors that could be applied to the skin for touch sensing, with some configurations supporting multitouch. Skinput~\cite{skinput} used a custom armband to detect bio-acoustic signals generated when a user tapped their host hand or arm. AudioTouch~\cite{audiotouch} used two active piezo elements placed on the back of the hand to detect palm touch gestures with a user-dependent model (89.3\% accuracy). SkinButtons~\cite{skinbuttons} used infrared sensors to detect touches to four locations around a watch-like device with an accuracy of 96.9\%. LumiWatch~\cite{lumiwatch} used an array of time-of-flight depth sensors to detect single finger touches with an accuracy of 99.3\%, but only within a very limited area on the forearm. 

\subsection{Instrumented Inputting and Receiving Arms}
Prior methods have also instrumented both the inputting arm and the input-receiving arm, which is perhaps the least practical configuration for consumers. The Sound of Touch~\cite{thesoundoftouch}, for instance, put ultrasonic piezo elements on both the inputting fingertip and receiving arm to detect a small set of points on the palm with an accuracy of 94.7\%. Touché~\cite{touche}, SkinTrack~\cite{skintrack}, and ActiTouch~\cite{actitouch} all use high-frequency AC signals to detect touches with accuracies of 84.0\%, 99.0\%, and 93.8\% respectively. While these prior methods show promising results, they all require both the inputting and receiving body locations to be instrumented.

\subsection{Touch Sensing with Cameras}
Researchers have also explored methods that avoid instrumenting a user's limbs and skin, chiefly by using cameras to remotely detect touches on surfaces, including the user's own skin. Early among these efforts was OmniTouch~\cite{omnitouch}, which used a shoulder-worn depth camera to track and detect finger touch events on the skin (and other surfaces in the environment), demonstrating a detection accuracy of 96.5\%. Note that OmniTouch could only reliably distinguish hover states when the finger was more than 4cm away from the skin, chiefly due to noise in the depth camera technology used at the time. Imaginary Phone~\cite{imaginaryphone} employed a head-mounted depth camera and detected touches using a series of heuristics, similar to OmniTouch. WatchSense~\cite{watchsense} mounted a structured-light-based depth camera on the user's forearm and reported a finger detection accuracy of 77.0\%. DIRECT~\cite{DIRECT} used a ceiling-mounted depth (and infrared) camera to detect touches to furniture and walls with an accuracy of 99.3\% at distances of up to 1.6m. FarOut Touch~\cite{farouttouch} later extended this range to up to 3m, while maintaining an accuracy of 96.9\%. TapLight~\cite{StructuredLightSpeckle} used a headset-mounted camera and laser emitter to detect touches on rigid surfaces with an f1-score of 95.3\%. The approach is able to detect touch-down events, but relies on conventional camera-based depth estimates to detect touch-ups. 

An important commonality of the above systems,  other than WatchSense~\cite{watchsense}, is that the user's arms were not instrumented in any manner (i.e., bare hands input). We also note that all of the above systems used depth cameras --- a very useful signal stream for knowing if a finger is truly touching or merely close to a surface. While depth camera cost has fallen, they are still less common and numerous on modern VR/AR headsets than RGB cameras.

We note that RGB cameras are particularly compelling as a signal source because they exist on all modern AR/VR headsets (for pass-through) and tend to have a large field of view (often several cameras are composited together). However, there is comparatively little work that has explored the use of RGB cameras for ad hoc touch tracking on the skin. 

PlayAnywhere~\cite{PlayAnywhere} was an early interactive tabletop, bare-hands input system. Although it used an infrared illuminator and camera, instead of  RGB cameras and natural light, some fundamental ideas are similar, including taking advantage of shadow shape to detect touches vs. hover. It is also similar in concept to the later ShadowTouch~\cite{shadowtouch} system, which used a head- or desk-mounted RGB camera, along with an LED wristband, to accentuate shadows cast by the fingertips onto the input surface. Palm+Act~\cite{palm+act} used a fixed RGB camera and optical flow to estimate finger pressure on the palm, but to our knowledge did not run an evaluation. TriPad~\cite{dupre:hal-04497640tripad} showed that hand tracking by itself could be used to make any surface temporarily touch sensitive by using the hand tracking 3D keypoints and dwells to determine surface locations. They reported an accuracy of 99.0\% for click detection on user-defined surfaces, but did not explore using the skin as a surface (likely because it moves and has variable and non-planar geometry). PressureVision++~\cite{PressureVisionPlusPlus}, extending the earlier PressureVision~\cite{PressureVision}, described a deep learning model for estimating fingertip contact on a variety of surfaces (but not the skin) using a desk-mounted RGB camera with an accuracy of 89.3\%. PressureVision++ estimated per-finger contact pressure on a nine-point scale with an IoU of 27.5\%. Inspired by this work, we also capture and estimate finger press force as an added dimension of touch input. 

\subsection{Contribution}
\autoref{tab:touch-systems-overview} summarizes key differences between closely related, vision-based, ad hoc touch tracking systems and  EgoTouch. While most systems report high accuracy, they often conduct their studies in static environments, with fixed cameras and lighting conditions. Moreover, most systems do not support input to the skin, and instead only support fixed and rigid surfaces like walls and tables. Further, few systems are evaluated without per-user or per-session calibration/training, nor include effects encountered on-the-go, including varying lighting and motion blur from being in motion. Overall, EgoTouch makes the following contributions:

\begin{itemize}
    \item A new and practical method for detecting finger touch events on the skin using conventional RGB cameras, like those already found in modern AR/VR headsets.
    \item Calibration-free model that generalizes across skin tone, hair density, input finger used, input location, and lighting condition.
    \item Real-time performance on mobile hardware.
    \item Exposes rich input metadata beyond binary touch (including finger force, pitch, and yaw), as well as standard UI touch events (touch down, touch up, and hover). 
\end{itemize}

\section{System}
\begin{figure*}[h]
    \centering
    \includegraphics[width=\linewidth]{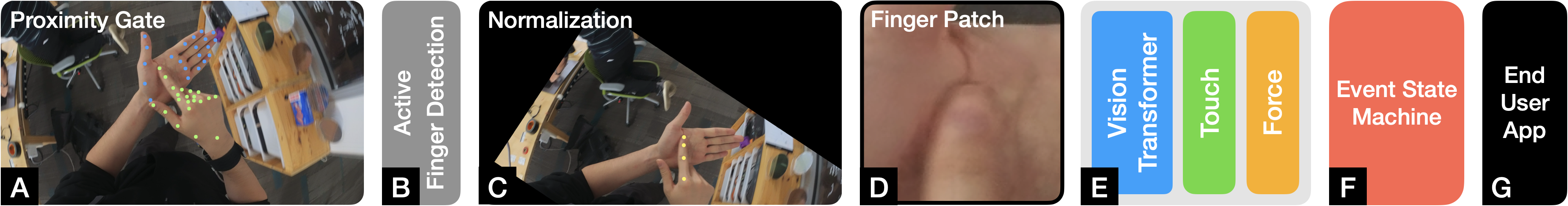}
    \caption{Software pipeline overview. First, hand keypoints are extracted from the RGB camera view (A; shown in blue and green for left and right hands). If the hands are sufficiently close together, our system then checks for fingers that might be indented for input (B). For each potentially active finger, the input image is rotated and scaled into a normalized form (C). From this, a roughly 4$\times$4~cm patch (with a resolution of 100$\times$100~px) is extracted around the fingertip (D) and fed into our touch and force estimation model (E). Model inferences drive a per-finger touch event state machine (F), which controls end-user applications (G).}
    \label{fig:pipeline}
\end{figure*}

We now describe the main components of our software pipeline (\autoref{fig:pipeline}), along with a brief description of our proof-of-concept hardware, which we used to make a functioning prototype. 

\subsection{Proof-of-Concept Hardware}
The Meta Quest 3 and Apple Vision Pro contain an array of RGB cameras that are composited together to provide high-resolution passthrough, hand tracking, and other interactive features. At the time of writing, the RGB streams were not accessible to developers and were only available and utilized inside Meta and Apple's proprietary software. For this reason alone, we affixed a wide-angle camera (ELP USBGS1200P01-L21, 1080p resolution) to our test headsets simply as a proof of concept. Furthermore, neither device provided support for external USB cameras, so we instead connected the camera to an M2 MacBook Air (2022), which was carried in a backpack during testing so the user could be untethered and mobile. The laptop performed all computations and forwarded any input events to the Quest 3 or Vision Pro over WiFi. Obviously, in a commercial implementation with low-level OEM access, the software and hardware would all be tightly integrated, likely as an additional feature of the hand-tracking pipeline. In our testing, we found the model to be lightweight enough to run in real-time on these devices (\autoref{sec:speed}).

\subsection{Proximity Gate \& Active Finger Detection}
The first step of our pipeline is to detect whether two hands/arms are present in front of the user. For this, we run Google's MediaPipe Hand model~\cite{zhang2020mediapipe}, which first detects hand bounding boxes, and then for present hands, proceeds to label 21 3D keypoints (\autoref{fig:pipeline}A). We then perform a quick check: are the hands close enough for an on-body touch interaction to even occur? (Specifically, we use within 3 palm lengths.) If so, we perform a second check: what, if any, fingers are pointing outwards (specifically, away from the wrist keypoint) and not tucked in? Any potentially "active fingers" are passed to the next step of our pipeline. 


\subsection{Normalization \& Finger Patch Extraction}
\label{sec:finger-path-extraction}
For each active finger found, we rotate the camera image to be finger-up aligned using the aforementioned 3D joint data. We also scale the image based on the distance between the distal interphalangeal (DIP) and fingertip joints of the finger, which helps to normalize the input image across users' different sized hands, and also across different operational distances. Example output from this process is shown in ~\autoref{fig:pipeline}C. We then take a roughly 4$\times$4~cm patch (with a resolution of 100$\times$100~px) centered at the fingertip. This process is completed for each active finger, allowing for multitouch tracking. We pass all active finger patches in parallel to our touch and force model, described next. 

\subsection{Touch \& Force Model}
Our model takes in a finger patch as input and outputs a touch prediction and press force estimate. This is done for all active fingers in parallel. Our model is a hybrid vision transformer model (\autoref{fig:pipeline}D) built on top of the FastViT T8~\cite{vasufastvit2023} backbone. This model encodes the image patches (in parallel if multiple) and produces image embeddings which are linearly transformed to embeddings of dimension 126. This is then concatenated with the $R$ and $\theta$ polar coordinates of the right fingertip relative to the left wrist (for right-handed input; would be swapped for left-handed input), on the axis formed between the wrist and the base (MCP) of the middle finger, forming a frame embedding of size 128. These embeddings are ReLU activated and linearly transformed into touch classification logits and force outputs.
To suppress single-frame errant output, we apply a three-frame median filter to both the touch and force outputs.
The model has 4.1 million trainable parameters. After training, we structurally reparameterize the model~\cite{vasufastvit2023} to a mathematically equivalent model with fewer branches and parameters. Our final model has 3.8 million parameters for inference.

\subsection{Model Training}
In our subsequent user study, we employ a leave-one-participant-out cross-validation scheme to train and test our models (i.e., a participant's data never appears in their training data and there is no calibration data or equivalent). We developed and trained our models using the PyTorch and PyTorch Lightning deep learning frameworks. The FastViT backbone is initialized with pre-trained ImageNet weights. For touch prediction, we compute a weighted binary cross-entropy loss, weighted by the ratio of touch and non-touch frames in the batch. For press force prediction, we compute a mean-squared error (MSE) loss and regress to the ground truth. The total loss of our model is a weighted sum of the touch and force losses (touch loss = 1$\times$, force loss = 5$\times$). Our model is trained end-to-end for eight epochs using the Adam optimizer, batch size of 128, and a learning rate of $0.0003$. Our models took about four hours to train on an Nvidia Titan V GPU. 

\subsection{Rich Input Metadata}
For each active finger, we include an array of rich input metadata on which sophisticated touch interactions can be built, exceeding that of contemporary touchscreens. As already discussed, our model estimates touch force (see, e.g.,~\cite{PressureVisionPlusPlus, pressurebasedkeyboard} for example uses). 3D keypoint data from our hand pose tracking phase provides a wealth of information, including finger identification (thumb, index finger, etc.) and 3D finger angle (see, e.g.,~\cite{xiao2015estimating} for uses in touch input). 

\subsection{Finger Touch State Machine}
Although our touch pipeline works on a frame-by-frame basis, we derive touch-down and touch-up events to expose a conventional touch input state machine. This is how a vast majority of user interfaces are driven, and so it is useful to make our approach immediately compatible with existing software. Our state machine also lets us trigger more specialized touch event handlers, such as onDrag and onLongClick that are available in e.g., Android~\cite{AndroidView2024}. 

\subsection{Performance}
\label{sec:speed}
Many computer vision models require large and expensive desktop-grade GPUs, making them impossible to run locally on today's mobile devices such as an AR/VR headset. An important goal of this work was to develop a model that was capable of running on mobile hardware as a background process. To assess this, we benchmarked our model on an Apple iPhone 12 Pro. We found that our model had an inference time of 0.75~ms when converted to CoreML. On a MacBook Air with M2 processor --- very similar to the hardware found in the Vision Pro --- our model had an inference time of 0.51~ms. In both cases, it means that we can run our model at 90 FPS or more (i.e., max framerate of the cameras) and consume a small fraction of the mobile device's processing power. 

We can compare this result to the best performing vision-based touch estimation model, PressureVision++ ~\cite{PressureVisionPlusPlus}, which has an inference time of 67~ms on a NVIDIA Titan V GPU (which we note is roughly the size of an AR/VR headset and consumes hundreds of watts of wall power).  

We note that the main bottleneck of our current prototype system is hand tracking, on which we rely on Google's MediaPipe hand model. On an Apple M2 processor, MediaPipe runs on the CPU at 76.4~FPS. Our full pipeline can only run as fast as our slowest component. In future work, we plan to create a fast hand-tracking model to remove this bottleneck. 

\section{User Study}
We now describe the user study we designed and executed to evaluate the performance of EgoTouch. 

\begin{figure}[b]
    \centering
    \includegraphics[width=\linewidth]{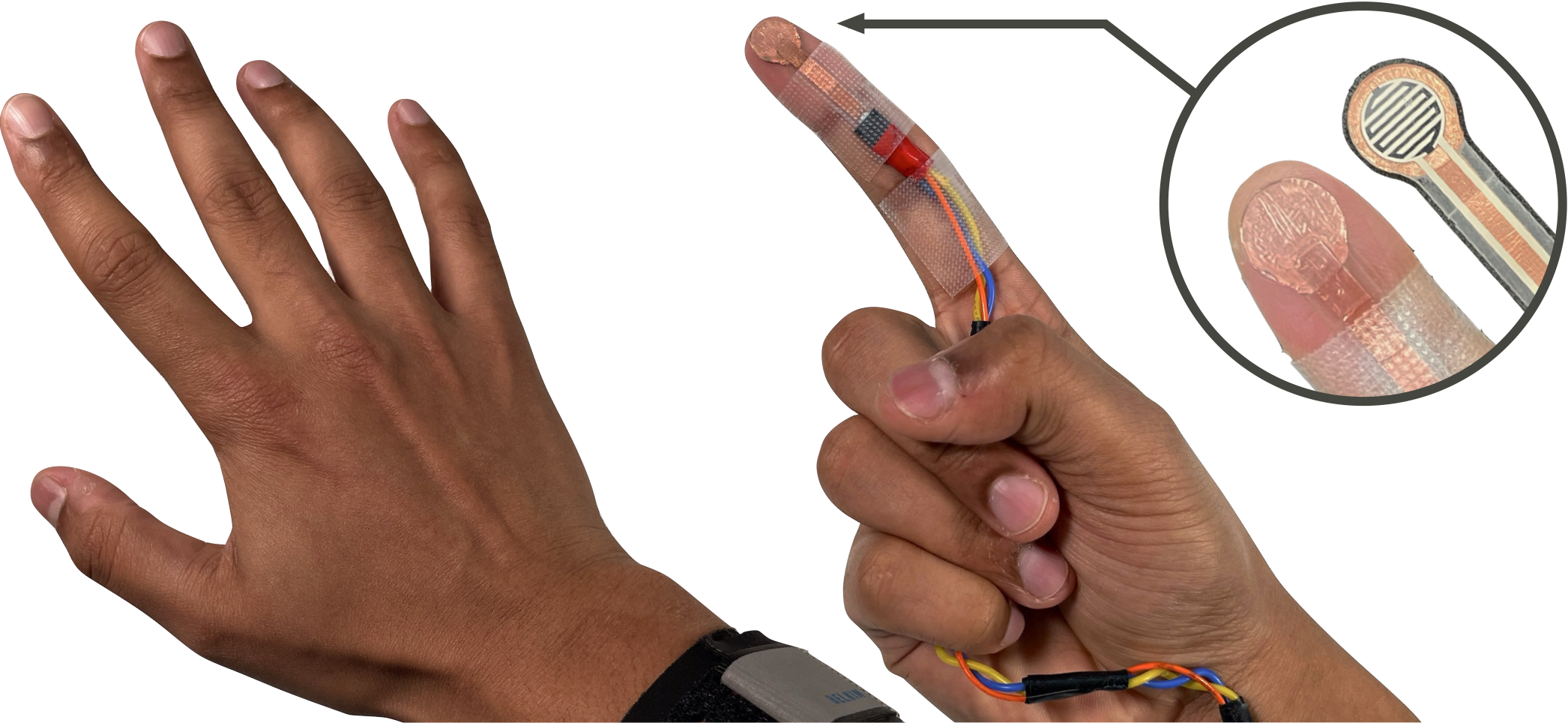}
    \caption{Our ground truth sensor (close-up in upper-right) is a force sensing resistor (FSR) overlaid with copper tape, which captures continuous press force and binary touch values, respectively. We attached this sensor to the underside of participants' fingers with skin tape, hiding it from the camera's view (left).}
    \label{fig:data collection apparatus}
\end{figure}

\begin{figure*}[h]
    \centering
    \includegraphics[width=\linewidth]{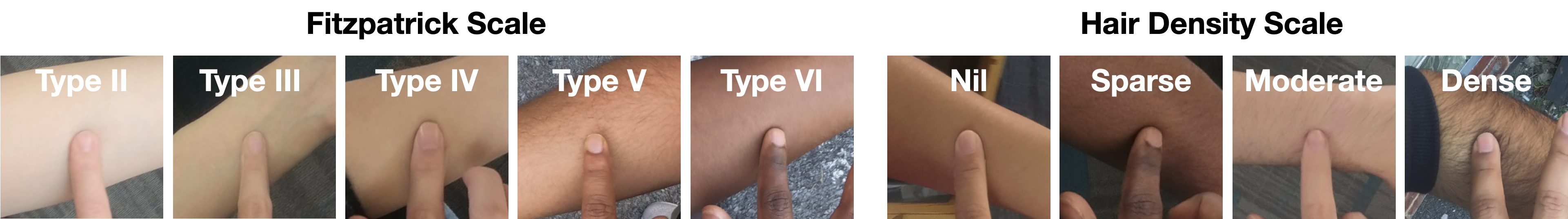}
    \caption{Example skin tones and hair densities collected in our user study. Scales based on~\cite{Fitzpatrick1975, armhairdensity}.}
    \label{fig:skintonehairdensity}
\end{figure*}

\subsection{Data Collection Apparatus}
Capturing ground truth data for computer vision models can be challenging because the use of tracking markers (e.g., retroreflectors, ArUco markers) or worn sensors (e.g., IMUs) can corrupt the visual data. For this reason, prior work such as PressureVision++~\cite{PressureVisionPlusPlus} had to rely primarily on weakly labeled data, which is useful, but less accurate. Other prior works --- especially systems using depth cameras for finger tracking (e.g., OmniTouch~\cite{omnitouch}) --- have struggled to disambiguate a finger hovering just above the skin vs. actually making contact. Given our goal was to overcome this problem and provide a touchscreen-like user experience, we needed a truly accurate ground truth. 

For this, we created a custom ground-truthing sensor (\autoref{fig:data collection apparatus}) invisible to the headset camera(s). Specifically, we used a small flexible force sensing resistor (FSR) overlaid with copper foil for binary capacitive touch sensing (using an MPR121 chip for capacitive readings). This is affixed to the underside of a user's finger during data collection using skin tape. Our FSR (p.n. DF9-40) was sensitive from 70g to 20kg of press force. We sampled both sensors at camera framerate using an ESP32-based QtPy board, applied a three-frame median filter to remove any single frame outliers, and transmitted data to a Macbook Air M2 (2022) laptop over USB serial. 

As noted previously, at the time of writing, neither Meta nor Apple provided an API to access the internal RGB camera streams of their headsets. As a proxy for these camera streams, we used an Insta360 One R camera with a wide-angle lens connected over USB to a laptop, capturing 1080p frames at 30~FPS. This camera was mounted to a user-adjustable head strap, tilted downwards to capture the spatial volume where arms typically interact in AR/VR. A laptop was placed in a backpack worn by participants so that they could be untethered and walk around during the study. 

\subsection{Data Synchronization}
\label{DataSynch}
Our RGB cameras and ground truth sensor data were not perfectly synchronized (due to different device clocks, USB contention, multi-threaded event handling, etc.). Additionally, the latency was not consistent --- we observed offsets up to ±133~ms (i.e., ±4 frames at 30~FPS). \autoref{fig:effectofdtw} offers a typical example of data we collected. Some prior work~\cite{StructuredLightSpeckle, omnitouch} corrected for this issue by applying some method of a windowed correction, allowing for some tolerance in the timing of events. In \autoref{fig:effectofdtw} (bottom), you can see the effect of a Dynamic Time Warping (DTW) post-processing (max ±4 camera frames), which corrects for synchronization errors but does not correct incorrect touch events. 

While we do report DTW-corrected results in Section~\ref{comparison-with-prior-work} (where we attempt to compare to prior work in an "apples-to-apples" fashion), we do not report this as our main result in Section~\ref{results}. Instead, as a more conservative estimate, we applied a one-time, post hoc, global offset of +33.3~ms (i.e., 1 frame forward at 30~fps) to our ground truth sensor data to better align it to our camera data. We note that in a real commercial system, there would be no need to align these two streams as there would be no ground truth sensor. 

\subsection{Participants}
We recruited 15 participants for our study (9 male, 6 female, mean age 26.4, all right-handed). Participants self-identified their skin tone on the Fitzpatrick scale~\cite{Fitzpatrick1975}, as well as their arm hair density~\cite{armhairdensity}, using two printed scales for reference. In terms of skin complexion, our participants had the following breakdown: 3 participants as type II, 8 as III, 1 as IV, 2 as V, and 1 as VI. In terms of arm hair density, our participants had the following breakdown: 4 participants as Nil, 6 as Sparse, 2 as Moderate, and 3 as Dense. Example skin tones and hair densities from our participants are shown in Figure~\ref{fig:skintonehairdensity}, with their permission.


\subsection{Study Procedure}
\label{sec:userstudy}

We designed our study to capture a wide variety of data for later analysis, including robustness across skin tones, hair densities, lighting conditions, indoor vs. outdoor locations, touch types, inputting finger, near hover vs. touch, and on-body touch location. To capture this variety of data, we divided our study into a series of small tasks.

After signing a consent form, participants were asked to self-identify their skin tone and arm hair density (see section above). We then equipped participants with a head-mounted camera connected via USB to a laptop in a backpack. Next, we cleaned the underside of the participant's index fingertip with rubbing alcohol and affixed our ground-truthing sensor with double-sided skin tape \autoref{fig:data collection apparatus}. Clear 3M Micropore tape was also added for additional support as needed. A velcro wrist strap was used to secure the QtPy board, which was also connected to the aforementioned laptop via USB. The experimenter manually set a threshold for the ground truth capacitance sensor such that the laptop would play a beep sound whenever a ground truth touch event was detected. 

We collected data from four different locations on the arm: inner forearm, inner palm, outer forearm, and back of the hand. At each of these locations, participants were instructed to perform four touch types: momentary \textit{taps}, long \textit{light presses}, long \textit{hard presses}, and finger \textit{hovers}. We also instructed participants to look at the location they were touching to allow for both hands to be visible in the camera view. To get natural data, we prompted participants to imagine various interfaces on their skin while performing touch events. For quick taps, we asked participants to imagine dialing a phone number on a touch screen. For light presses, we asked users to imagine gently painting dots on their skin and holding for a moment, with a light amount of force, before moving their finger to a new location. Hard presses were similar, but participants were asked to apply more force. 
Finally, for finger hovers, participants were instructed to repeatedly move their finger as close to their skin as they could without touching their skin. Any accidental touches (captured by our ground truth sensor) were removed from the captured data. 

The above procedure was repeated twice, once at an indoor location and once outdoors. We used a Sper Scientific 840022 Advanced Light Meter to record illumination levels, which ranged from 12 Lux outside at night with distant street lights to 35,000 Lux outdoors in full sun. 

All of the above data was collected with participants using their right index finger, as this is the most popular digit employed for touch input. As a final stage, we collected a small sample of data from non-index fingers (which required removing and re-affixing our ground truth sensor). Specifically, participants were asked to perform hard presses on their inner palm with their thumb, middle, ring, and pinky fingers, adding four more rounds of data.

In summary, each of our 15 participants completed 36 rounds of data collection (4 body locations $\times$ 4 touch types $\times$ 2 lighting locations + 4 non-index fingers). Each of these data collection rounds lasted approximately 30 seconds and yielded roughly 900 frames of data. In total, the study procedure took about 45 minutes, and paid \$20 in compensation.


We note that our study did not constrain participants' movement or environment. Participants sat down, stood still, and walked around at will, both indoors and outdoors, resulting in a purposely large variety of lighting conditions, head movements, and arm positions. Whenever possible, the study was conducted at the participant's location to introduce further variation. This is in contrast to many prior studies where participants sat or stood in stable environments with controlled lighting and fixed cameras (e.g., ~\cite{PressureVision,PressureVisionPlusPlus,StructuredLightSpeckle, shadowtouch, dupre:hal-04497640tripad, DIRECT, farouttouch, watchsense}); summarized in Table \ref{tab:touch-systems-overview}.

\subsection{Train-Test Scheme}
For each participant, we held out that participant's data for evaluation and combined the other 14 participants' data into a training set (all combinations with results combined; i.e., leave-one-participant-out cross-validation). Such a train-test scheme means there is no "calibration" or equivalent data in the training set, and it is if a general pre-trained model is seeing the user for the first time (i.e., "out-of-the-box" accuracy). Note also that our model only runs when it passes our hand proximity check. Rather than include these frames, where our touch accuracy is 100\% (i.e., never touching), we only consider frames and report accuracies when our model actually runs (and there may be some degree of ambiguity).

During training, we applied various data augmentations to extracted finger patches. Specifically, the brightness, contrast, saturation, and hue of the patches were randomly jittered by factors of 0.2 to 2.2, 0.8 to 2.0, 0.5 to 1.5, and -0.1 to 0.1, respectively. Random patches were also flipped horizontally. 
Unlike other image models, we did not apply scale or rotation augmentations since our finger patch extraction pipeline already aligned and scaled the images to the finger (see Section~\ref{sec:finger-path-extraction} and Figure~\ref{fig:skintonehairdensity}).

\section{Results and Discussion}
\label{results}
We now discuss our main study findings, reporting results as frame-level accuracies unless otherwise noted. In Section \ref{comparison-with-prior-work}, we additionally report accuracy using a small-window dynamic latency correction (see also Section \ref{DataSynch}) to better compare to prior work that used this method. 

\subsection{Effect of Skin Tone \& Hair Density}
\begin{figure}[t]
    \centering
    \includegraphics[width=\linewidth]{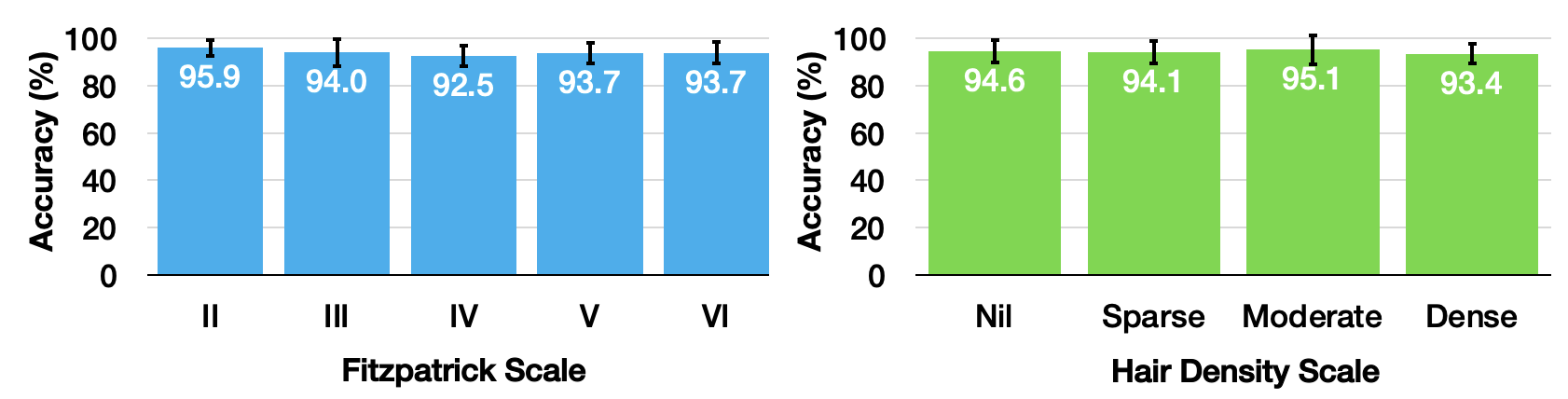}
    \caption{Frame-wise touch classification accuracy vs. skin tone (left) and hair density (right).}
    \label{fig:skin-hair}
\end{figure}

\begin{table}[b]
    \centering
    \begin{tabularx}{.8\linewidth}{Xd{3.3}d{3.3}}
        \toprule    
        \textbf{Factor} & \multicolumn{1}{c}{\textbf{BF$_{10}$}} & \multicolumn{1}{c}{\textbf{Error (\%)}}\\
        \midrule
        Skin Tone & 0.169 & 0.430 \\
        Hair Density & 0.138 & 0.476 \\
        \midrule
        Touch Location & 1.023 & 0.582 \\
        Lighting & 1.025 & 0.023 \\
        Touch Type & 4.277 & 0.694 \\
        \bottomrule
    \end{tabularx}
    \vspace*{2mm}\caption{Results of a Bayesian factor analysis.}
    \label{tab:stats}
\end{table}
We first conducted a between-subject Bayesian factor analysis on all our data (across 4 touch locations, 4 touch-types, and 2 location conditions) to investigate the likelihood of either skin tone or hair density affecting our accuracy. The analysis showed moderate evidence ($BF_{10} < .33$) that accuracy was not affected by these two factors; see \autoref{tab:stats} and \autoref{fig:skin-hair}. This is an encouraging result, as this can be a major issue in some human-sensing computer vision systems. Thus, results are reported across all of our participants in subsequent analyses. 


\subsection{Touch Accuracy}
To evaluate our model's touch classification accuracy, we used our 32 rounds of index finger data (across body input locations, touch types, lighting conditions, etc.) which contained 125,656 touching frames and 181,844 non-touching/hovering frames. Our mean frame-wise classification accuracy was 94.9\% (std=3.5). True positive touch detection rate was 96.4\% and false positive rate was 5.6\%. In Section \ref{comparison-with-prior-work} we report frame-wise accuracy with a small latency tolerance to account for asynchrony between our ground truth and camera streams (see also Figure \ref{fig:effectofdtw}). 


Rather than consider accuracy on a frame-by-frame basis, we can also look at event-wise "click" accuracy (i.e., touch-down and touch-up events). Across the 2783 touch events captured in our user study, we found a mean click event accuracy of 95.6\% (std=8.3). Note that in order for an event frame to be considered correct, the predicted event and ground truth event had to occur in the exact same frame (at our study capture rate of 30 FPS). Note this is different from some prior work that used a time tolerance of a few frames when computing event-level accuracy.



\subsection{Force Estimation}
Also using our index finger data, we evaluated our model's force estimation accuracy. Forces in our study ranged from 0~N to 3.5~N (mean=1.1~N). From the literature~\cite{1Nref1, 1Nref2}, forces exerted on a typical touch screen are roughly 1~N. Our model estimated force with a mean absolute error of 6.8\% (std=3.9) across all participants. Put simply, if a user applied a ground truth force of 1.0N, a 6.8\% error would be ±0.068N of force. If we separate our dataset into two force ranges: [0, 1N) (soft and medium presses), and [1N, 3.5N) (hard presses), classification accuracy is 97.9\%. Such touch metadata could be used to trigger functionality like that seen on iPhone models with "3D Touch" / "Force Touch". 



\subsection{Effect of Touch Location}

\begin{figure}[b]
    \centering
    \includegraphics[width=0.80\linewidth]{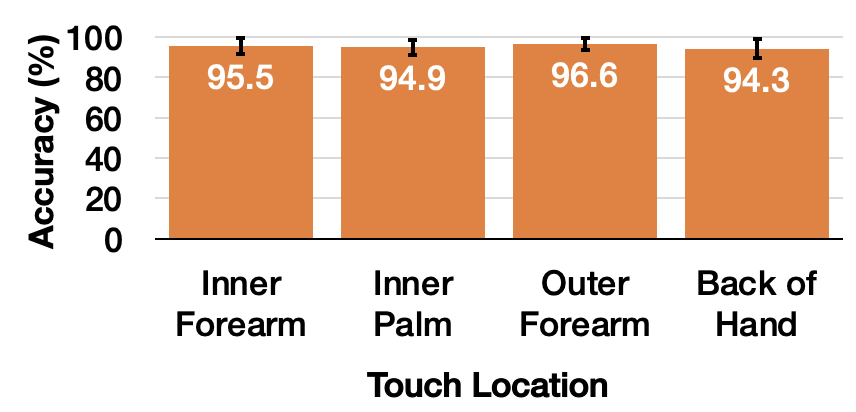}
    \caption{Frame-wise touch classification accuracy vs. touch location.}
    \label{fig:location}
\end{figure}

We conducted a Bayesian factor analysis to investigate the likelihood that Touch Location affected the accuracy of touch classification. The analysis of anecdotal evidence ($BF_{10} < 1$) shows that the accuracy is not affected by the Touch Location (\autoref{tab:stats} and \autoref{fig:location}). In particular, we notice that the outer forearm has anecdotal evidence to be higher than the inner-palm ($BF_{10}=1.855$) and the back of the hand ($BF_{10}=3.225$). 


\subsection{Effect of Lighting}

\begin{figure}[b]
    \centering
    \includegraphics[width=1.0\linewidth]{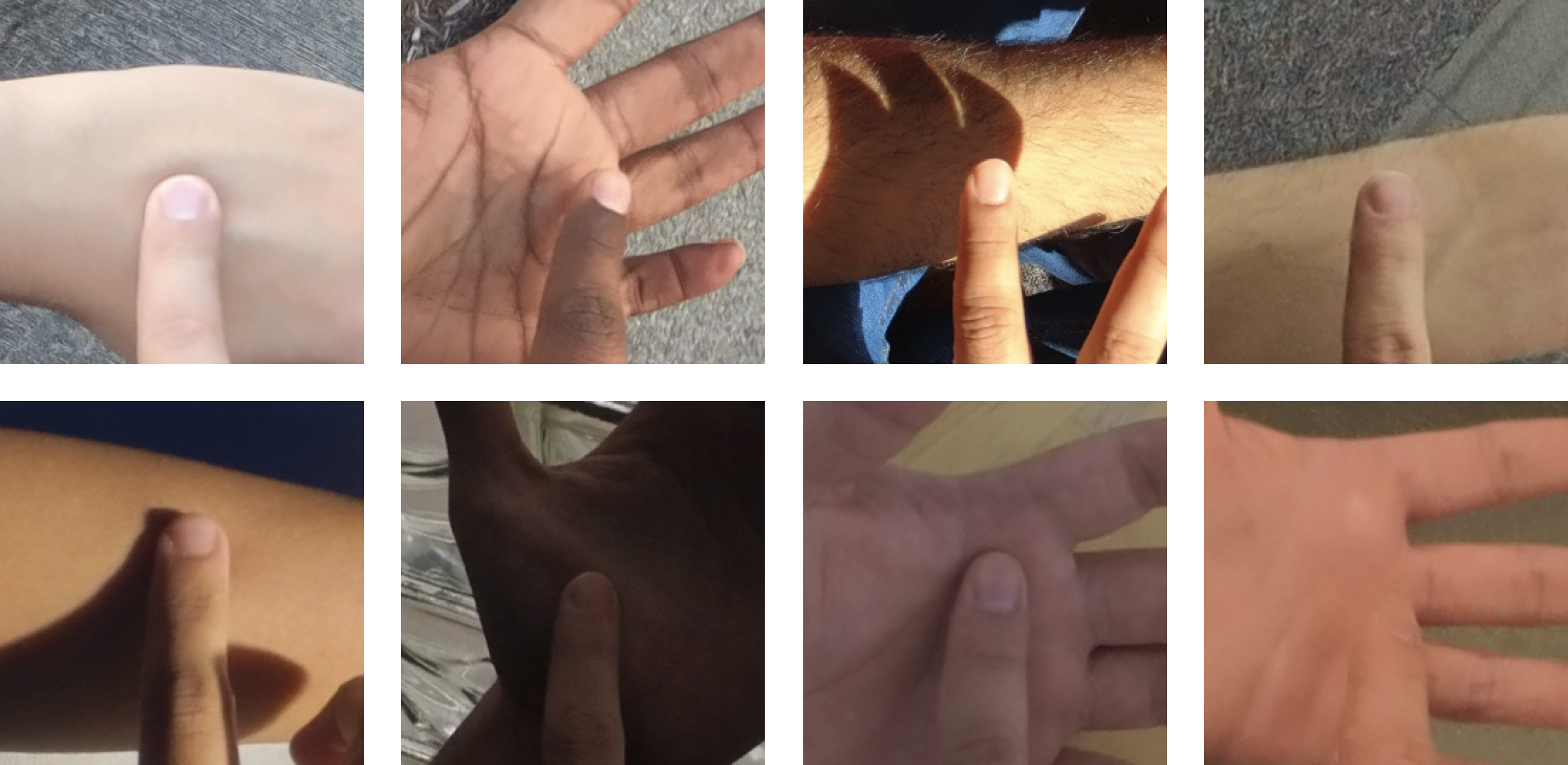}
    \caption{Our study included diverse lighting conditions, including indoor and outdoor settings, diffuse and direct light, and illumination from artificial lights and the Sun. }
    \label{fig:lighting-conditions}
\end{figure}

Our study included indoor and outdoor data collection conditions, under both artificial and sun light, and even at night for two participants (example study frames provided in \autoref{fig:lighting-conditions}). The ambient illumination varied from 12 Lux (outside at night with distant metal-halide street lights), through 160 Lux (indoors in a dimly lit office using florescent tubes), all the way to 35,000 Lux (skin in direct sunlight). 
A Bayesian paired samples t-test did not suggest lighting condition impacted accuracy with anecdotal evidence (\autoref{tab:stats}). Mean touch classification accuracy was 94.7\% (std=4.2) indoors and 95.7\% (std=3.9) outdoors.

This result was somewhat surprising, as we anticipated low-light operation to be a weak spot of our approach. However, it seems the camera sensor was able to capture sufficient detail even at low light levels, and the noise and motion blur present in the image can be handled by our model. At the other end of the spectrum, in very bright conditions when there is little sensor noise or motion blur (due to short camera shutter speeds), the model is sufficiently generalized to handle different shadow types, including harsh oblique shadows from direct light sources. 

\subsection{Effect of Touch Type}

We also conducted a Bayesian factor analysis to investigate the likelihood that Touch Type affected performance. The analysis showed evidence ($BF_{10} > 3$) that accuracy is affected by Touch Type (see \autoref{tab:stats}). In particular, using post hoc comparisons, we observed that hard presses have better accuracy than light presses ($BF_{10} = 25.91$). Taps and light presses showed equal performance ($BF_{10} = .85$) as did taps and hard presses ($BF_{10} = .48$). 

Hard press events had the highest accuracy at 97.7\% (std=4.5). This is not surprising as they produce the most exaggerated deformations of the host skin. Light presses are harder to observe, especially on the back of the hand, which results in a decreased accuracy of 95.2\% (std=11.6). We note that in our light press trials, several participants remarked this was lighter than they would press on a conventional touch screen. Tap accuracy was slightly lower at 94.5\% (std=15.4), likely due to the momentary nature of the interaction and motion blur from the rapid action. 

Finally, as described in \autoref{sec:userstudy}, we also asked our participants to hover their finger as close as they could above their skin. This condition was purposely included to assess robustness to close hovers, a challenge for prior depth-camera-driven touch systems. For these hovers, our model was able to correctly detect hovers (i.e., that they were not touching) with an accuracy of 98.6\% (std=1.9).




\subsection{Effect of Finger Type}
\begin{figure}[t]
    \centering
    \includegraphics[width=\linewidth]{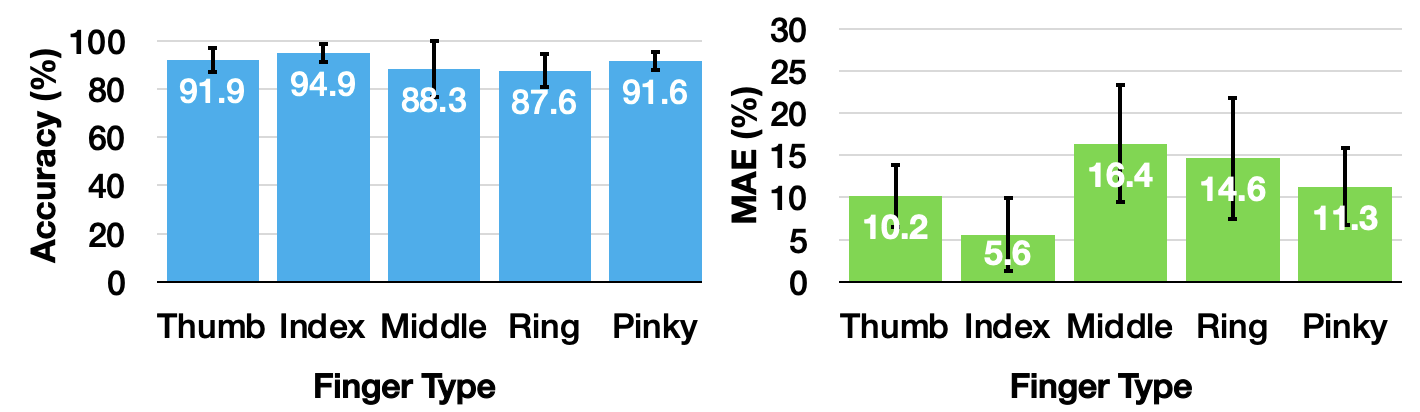}
    \caption{Frame-wise touch classification accuracy (A) and force estimation error (B) for the five fingers. Note that the index finger performs better due to significantly more training data; there is no reason to believe similar performance cannot be achieved for the other fingers.}
    \label{fig:acc_vs_finger_type}
\end{figure}

While our study mainly focused on the use of the index finger for input, we also collected one round of data for each of the other four fingers. 
This means that each of the non-index fingers makes up only about 2.8\% (1 out of 36 rounds) of our training data, and thus this result should only be considered preliminary. Overall, frame-wise touch classification accuracy was 91.9\%, 88.3\%, 87.6\%, and 91.6\% for the thumb, middle, ring, and pinky fingers, respectively (Figure \ref{fig:acc_vs_finger_type}, left). For touch force estimation, frame-wise mean absolute error was 10.2\%, 16.4\%, 14.6\%, and 11.3\% for the thumb, middle, ring, and pinky fingers, respectively (Figure \ref{fig:acc_vs_finger_type}, right).

\subsection{Comparison to Prior Work}
\label{comparison-with-prior-work}
\begin{figure*}[t]
    \centering
    \includegraphics[width=\textwidth]{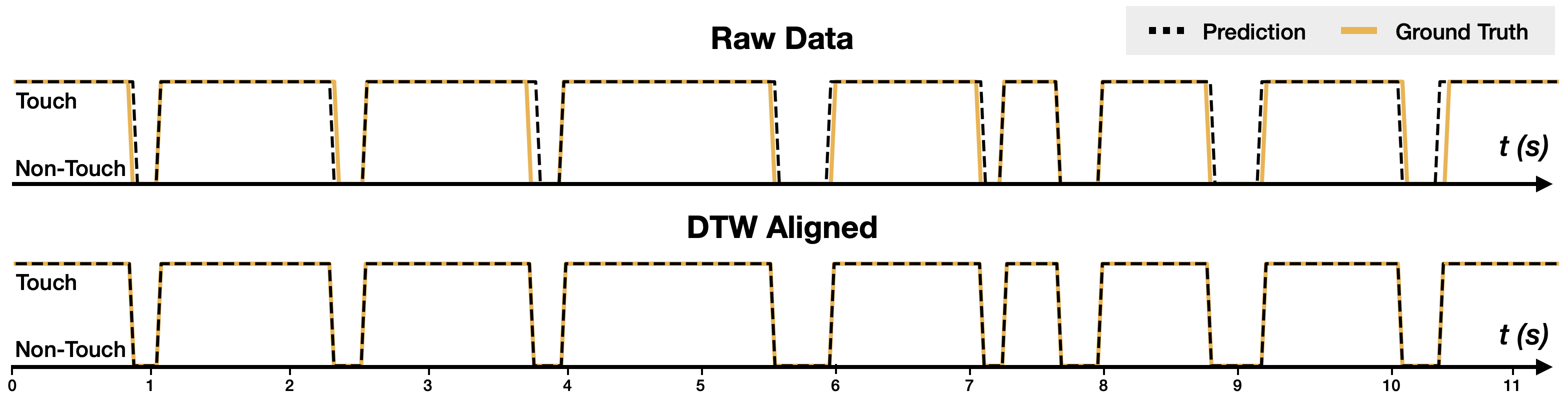}
    \caption{A real-world, ten-second sequence from our user study containing eight long press events of varying durations. Yellow is ground truth (captured by our worn capacitive sensor) while the black dotted line is our model's touch prediction. To remove the confound of varying signal alignment (due to unsynchronized sensor/camera and variable USB event dispatching latency), we can apply a small-window dynamic time warping correction (max ±4 frames, which is ±133ms). Before alignment, the frame-wise accuracy of the raw sequence is 96.3\%, even though every touch event is accurately detected. After alignment, frame-wise accuracy is 100\%. In this paper, for transparency, we report both fixed-latency frame-wise accuracy, as well as variable-latency-corrected frame-wise accuracy (reported in Comparison to Prior Work, Section \ref{comparison-with-prior-work}). }
    \label{fig:effectofdtw}
\end{figure*}

Although our recording pipeline used millisecond-accurate system timestamps to synchronize our data streams, we had little control lower down the stack. Notably, our camera and ground truth touch sensor were not synchronized (i.e., run on their own clocks). Compounding this further, the latency in USB event handling on our laptop varied as well. Figure~\ref{fig:effectofdtw} provides an example ten-second sequence from our study, containing eight touch events. If we naively compute frame-wise accuracy, the top sequence is 96.3\% accurate, even though every touch event was accurately detected. 

To remove the confound of varying latency due to our data collection apparatus, and produce a more true-to-life result (i.e., a button clicked is a button clicked), we can apply a small-window dynamic time warping correction (up to ±4 camera frames, which equates to ±133ms). This helps to align the data, but not so far as to correct for errant behavior. After this small alignment correction, the bottom signal in Figure \ref{fig:effectofdtw} has a frame-wise accuracy of 100\%. 

If we apply this same alignment correction to our full study data, our system's frame-wise touch classification accuracy increases from 94.9\% to 97.3\%. We believe the latter number is a more faithful representation of our system's real-world performance, as a commercial system would not have synchronization issues (i.e., there would be no ground truth sensor, just a camera stream). We note that this type of post hoc alignment of ground truth and prediction data has been used in other similar work (e.g., TapLight~\cite{StructuredLightSpeckle} and OmniTouch~\cite{omnitouch} used ±300ms and ±500ms event correction windows respectively).

From a technical standpoint, the most related prior work to our own is PressureVision++~\cite{PressureVisionPlusPlus}. This work does not directly estimate touches but rather predicts one of nine pressure levels (any of which above zero can be considered a touch). We ran the publicly available PressureVision++ model on several clean and flat desks, but we found performance to be inconsistent (perhaps because the model was trained with a specific and static setup). We also ran our study data through the model, but performance was very poor (not surprising as it was never intended for skin input). 





\section{Limitations and Future Work}
EgoTouch's touch detection and force estimation performance is good, but like any other machine learning system, it must be further tested to support numerous edge cases before it can be ready for commercial adoption. We strove to make our study as comprehensive as possible, but there are still important questions worth answering: How does nail polish, acrylic nails, jewelry, and other treatments and accessories affect performance? Will the system work with gloves? Can the approach scale to non-skin surfaces? Additional studies must also be conducted to fully evaluate the multitouch capabilities.

For our small user study, we endeavored to recruit a diverse set of 15 participants. Of course, this is not fully representative and future models will need to be trained on a much larger set of participants. Another experimental compromise was to not include drag movements as a touch type. We had planned to include drags, but found in pilot testing that the tape holding our ground truth sensor to the skin kept peeling up. In later pilots, we asked participants to “be careful to not peel off the sensor”, but this in turn modified their touch behavior artificially, and we ultimately decided to drop this input type from our study. We found, however, that our current model can reasonably support drags, as seen in the Video Figure.

We also note that EgoTouch was sometimes hindered by the poor hand detection and tracking accuracy of MediaPipe, especially when the hands were interacting and overlapped. We witnessed times when left hands would be mistaken for right hands and vice versa, and also cases where hands were not detected at all even when clearly visible. Although our pipeline attempts to filter out such erroneous events with basic heuristics, some poorly tracked data inevitably make it into our training and evaluation data sets, hurting performance. 
Additionally, as discussed in \autoref{sec:speed}, the main bottleneck of our system's performance was MediaPipe hand tracking (which is many times slower than our model). Training our own hand model was outside the scope of this work, but we now believe a faster/better hand model is key in fully unlocking EgoTouch, and we hope to explore this in future work.

Finally, using an always-on camera for touch detection can raise privacy concerns. Fortunately, our pipeline is able to run locally, such that video data never has to leave the device. In this way, it can operate like hand tracking and passthrough video does on today's AR/VR headsets, which also utilize always-on cameras. We also note that the camera we used to collect data was capped at 30 FPS at 1080p. In dimly lit environments, frames often had motion blur. Furthermore, 30 FPS was sometimes not fast enough to capture very short events such as quick taps. In later stages of the project, we moved to a USB webcam offering 90 FPS and noticed improved performance. 






\section{Conclusion}
In this paper, we presented EgoTouch, a system that uses RGB camera(s) --- like those already found in modern AR/VR headsets --- to detect finger-to-skin touches and estimate press force. Initial results suggest our approach can work across varying skin tones, arm hair densities, lighting conditions, touch locations, touch types, and different fingers. EgoTouch also exposes rich per-finger metadata, including press force and 3D finger angle, more fully unlocking the potential for on-skin interfaces.

\begin{acks}
We are particularly grateful to Prof. Sven Mayer from the Media Interaction Lab at LMU Munich for his help with data analysis. We also thank our participants for granting us permission to share photos captured during the study and their recorded data.
\end{acks}

\bibliographystyle{ACM-Reference-Format}
\bibliography{sample-base}
\end{document}